# Chaotic trigonometric Haar wavelet with focus on image encryption

**Sodeif Ahadpour[1], Yaser Sadra[1,2]**

[1]Faculty of Sciences, University of Mohaghegh Ardabili, Ardabil, Iran.

[2]Department of computer sciences, Shandiz Institute of Higher School, Mashhad, Iran.

Corresponding author e-mail: ahadpour@uma.ac.ir

## Abstract

In this paper, after reviewing the main points of Haar wavelet transform and chaotic trigonometric maps, we introduce a new perspective of Haar wavelet transform. The essential idea of the paper is given linearity properties of the scaling function of the Haar wavelet. With regard to applications of Haar wavelet transform in image processing, we introduce chaotic trigonometric Haar wavelet transform to encrypt the plain images. In addition, the encrypted images based on a proposed algorithm were made. To evaluate the security of the encrypted images, the key space analysis, the correlation coefficient analysis and differential attack were performed. Here, the chaotic trigonometric Haar wavelet transform tries to improve the problem of failure of encryption such as small key space and level of security.

## 1. Introduction

Recently, wavelet analysis is a strong mathematical tool for signal and image analysis [1, 2, 3, 4, 5, 6]. The wavelets have been applied in many fields such as image compression, data compression, encrypted data and a lot more [7, 8, 9, 10, 11]. The image processing and encrypting based on discrete transform is currently the processing technique [12, 13, 14, 15]. The image and data compression applications have expanded in the field of the multimedia which requires the high performance, speedy digital video and audio capabilities [16, 17, 18, 19]. One of the methods for image compression is based on Haar wavelet transforms, which are used in the JPEG–2000 standard as wavelet–based compression algorithms [19, 20, 21]. In this work, we first introduce a new perspective of Haar wavelet which can be used in many more fields. Then, a chaotic Haar wavelet is made based on a chaotic map, which can be using in image and data encryption. Experimental results confirm this topic. This paper is ordered as: An introduction of Haar wavelet and trigonometric chaotic map discusses in section 2. We introduce the new perspective Haar wavelet that can be used to image compression and encryption in section 3. Then, we present the encryption methods proposed in this research in section 4. Next, the experimental results and statistical analysis discusses in section 5 and finally, in Section 6, we conclude this work.

## 2. Background

In this section, we demonstrate basic concepts Haar wavelet and chaotic maps to make a new perspective Haar wavelet.

- Haar Wavelet

Wavelet analysis and Fourier analysis are similar each other in that they allow a purpose function over an interval to be demonstrated aspect of the orthonormal function basis. Haar wavelet is simplest possible wavelet; hence, it most widely used in the various sciences. Haar transform or Haar wavelet transform is the simplest of the wavelet transforms and the first known wavelet which was proposed in 1909 by Alfred

Haar [4]. Haar transforms cross-multiply a function versus the Haar wavelet with various shifts and stretches, like Fourier transforms cross-multiply a function versus a sine wave with two phases and many stretches. The Haar wavelet is constructed based on the multiresolution analysis, which is generated by the scaling function [4, 5]:

$$\phi(x) = \begin{cases} 1 & 0 \leq x < 1 \\ 0 & other\ wise \end{cases} \quad (1)$$

In result, the Haar wavelet is the function:

$$\psi(x) = \begin{cases} 1 & 0 \leq x < \frac{1}{2} \\ -1 & \frac{1}{2} \leq x < 1 \\ 0 & otherwise \end{cases} \quad (2)$$

Their graphs are given in Fig.1. In other words, with regard to a more comprehensive mathematical definition, the scaling function and wavelet function are as following [4, 5]:

$$\phi(x) = \sum_{l \in z} p_l\, \phi(2x - l) \quad (3)$$

or for any $j \in Z$, the set of functions

$$\{\phi_{j,l}(x) = 2^{\frac{j}{2}} \phi(2^j x - l); l \in Z\}$$

where, $l$ determines the position of $\phi_{j,l}(x)$ along the x-axis, $j$ determines $\phi_{j,l}(x)$'s width. Then

$$\psi(x) = \sum_{l \in z}(-1)^l\, \overline{p_{1-l}}\, \phi(2x - l) \quad (4)$$

that

$$p_l = 2 \int_{-\infty}^{\infty} \phi(x)\, \overline{\phi(2x-l)}\, dx \quad and \quad \widetilde{p}_l = 2^{\frac{-1}{2}} p_l.$$

The $\widetilde{p}_l$ coefficients in here are called scaling function coefficients. In result, the Haar scaling function and wavelet function are defined as ($p_0 = p_1 = 1$) [4, 5, 6]

$$\phi(x) = \phi(2x) + \phi(2x - 1)$$

$$\psi(x) = \phi(2x) - \phi(2x - 1)$$

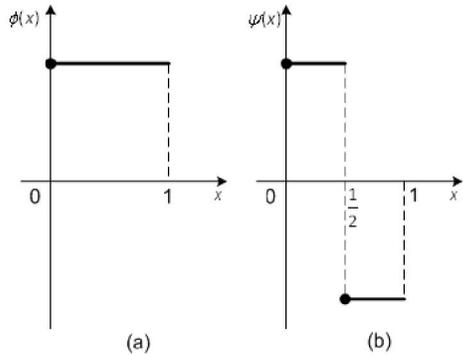

Fig.1. Graph of $\phi(x)$ and $\psi(x)$ of Haar wavelet.

Two-dimension Haar wavelet is created based on taking the tensor product of one-dimensional wavelets. As regards $\phi$ and $\psi$ are the scaling function and wavelet; we can be demonstrated the multiresolution analysis on a two-dimension signal $f_{n,n}$ as following [4, 6]:

$$f_{n,n} = LL_{n-1,n-1} + LH_{n-1,n-1} + HL_{n-1,n-1} + HH_{n-1,n-1}$$

Where

$LL_{n-1,n-1}$ is a linear combination of $\psi(2^{n-1}x - l)f_{n,n}(x,y)\psi(2^{n-1}y - \hat{l})$

$LH_{n-1,n-1}$ is a linear combination of $\psi(2^{n-1}x - l)f_{n,n}(x,y)\phi(2^{n-1}y - \hat{l})$

$HL_{n-1,n-1}$ is a linear combination of $\phi(2^{n-1}x - l)f_{n,n}(x,y)\psi(2^{n-1}y - \hat{l})$

$HH_{n-1,n-1}$ is a linear combination of $\phi(2^{n-1}x - l)f_{n,n}(x,y)\phi(2^{n-1}y - \hat{l})$

Where L is approximation sequence or low-pass filter in horizontal or vertical directions of the matrix and also, H is details sequence or high-pass filter in horizontal or vertical directions of the matrix [4, 6].

In the image analysis application, a two-dimension signal $f_{n,n}$ can be a plain image. The two-dimensional wavelet transform decomposes the plain image to four sub-images (LL, LH, HL and HH). In other words, these sub-images are filtered with low-pass (L) and high-pass filters (H). LL is the approximated plain image with quarter of the initial size that has been filtered using of low-pass filters in horizontal and vertical directions. HL is a details image which has been filtered using of high-pass filter in the vertical direction and low-pass filter in horizontal direction. LH is a details image which has been filtered using of low-pass filter in the vertical direction and high-pass filter in horizontal direction. HH is a details image which has been filtered using of high-pass filters in both horizontal and vertical directions [4, 6]. The LL image can be decomposed to four new sub-images which make a tree of sub-images as shown in Fig. 2.

In other words, the Haar transform can be presented in matrix form $F = HMH^T$ where M is a $n \times n$ plain matrix, H is a $n \times n$ Haar transform matrix, and F is the $n \times n$ resulting transform matrix. An example of a $4 \times 4$ Haar transform matrix is shown in Eq. (5) [6]:

$$H_{4\times4} = \begin{bmatrix} \frac{1}{2} & \frac{1}{2} & \frac{1}{2} & \frac{1}{2} \\ \frac{1}{2} & \frac{1}{2} & \frac{-1}{2} & \frac{-1}{2} \\ \frac{1}{\sqrt{2}} & \frac{-1}{\sqrt{2}} & 0 & 0 \\ 0 & 0 & \frac{1}{\sqrt{2}} & \frac{-1}{\sqrt{2}} \end{bmatrix} \quad (5)$$

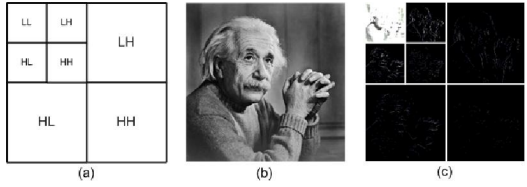

Fig.2. (a) Two dimensional 2-leveles Haar wavelet decomposition. (b) The plain image ''Einstein.bmp''. (c) Two levels Haar wavelet decomposition of the plain image ''Einstein.bmp''.

- The chaotic trigonometric maps:

In this section, we first review one-parameter chaotic maps that can be used as basis of chaotic trigonometric maps. The one-parameter chaotic maps are defined as the ratio of polynomials of degree N [22, 23, 24]:

$$\varphi_N^1(x, a) = \left(1 + (-1)^N F_1^2\left(-N, N, \frac{1}{2}, x\right)\right) \times \frac{a^2}{(a^2+1) + (a^2-1)(-1)^N F_1^2\left(-N, N, \frac{1}{2}, x\right)}$$

$$= \frac{a^2 (T_N(x^{\frac{1}{2}}))^2}{1 + (a^2-1)(T_N(x^{\frac{1}{2}}))^2}$$

and

$$\varphi_N^2(x, a) = \left(1 - (-1)^N F_1^2\left(-N, N, \frac{1}{2}, (1-x)\right)\right) \times \frac{a^2}{(a^2+1) - (a^2-1)(-1)^N F_1^2\left(-N, N, \frac{1}{2}, (1-x)\right)}$$

$$= \frac{a^2 (U_N((1-x)^{\frac{1}{2}}))^2}{1 - (a^2-1)(U_N((1-x)^{\frac{1}{2}}))^2}$$

where N is an integer greater than one. Also,

$$F_1^2\left(-N, N, \frac{1}{2}, x\right) = (-1)^N \cos\left(2N \, ArcCos\left(x^{\frac{1}{2}}\right)\right)$$
$$= (-1)^N T_{2N}(x^{\frac{1}{2}})$$

is the hyper geometric polynomials of degree N. $T_N(U_N(x))$ are chebyshev polynomials of type I (typeII), respectively. The chaotic trigonometric maps can be their conjugate maps which are defined as:

$$\widetilde{\varphi_N^1}(x, a) = h \, O \, \varphi_N^1(x, a) \, O \, h^{-1}$$
$$= \frac{1}{a^2} Tan^2\left(N \, ArcTan\left(x^{\frac{1}{2}}\right)\right),$$

$$\widetilde{\varphi_N^2}(x, a) = h \, O \, \varphi_N^2(x, a) \, O \, h^{-1}$$
$$= \frac{1}{a^2} Cot^2\left(N \, ArcTan\left(x^{\frac{-1}{2}}\right)\right),$$

Here, conjugacy means that the invertible map $h(x) = \frac{1-x}{x}$ maps $x = [0,1]$ into $[0, \infty)$ [22, 23, 24]. We denote the chaotic trigonometric maps $(\widetilde{\varphi_N^1}(x, a), \widetilde{\varphi_N^2}(x, a))$ with $(f_1(x, a), f_2(x, a))$ in order to simplify the calculation in this paper. Therefore, the chaotic trigonometric maps are as follows:

$$f_1(x_n, a_1) = \frac{1}{a_1^2} Tan^2\left(N_1 \, ArcTan\left(x_{n-1}^{\frac{1}{2}}\right)\right), \quad (6)$$

$$f_2(x_n, a_2) = \frac{1}{a_1^2} Cot^2\left(N_2 \, ArcTan\left(x_{n-1}^{\frac{-1}{2}}\right)\right), \quad (7)$$

Here, we use the chaotic trigonometric maps as generic symmetric non-linearly coupled maps which are the dynamical chaotic maps as following:

$$f_{coupled} = (1 - \varepsilon) f_1(x) + \varepsilon f_2(x)$$

In other words,

$$x_{n+1} = (1 - \varepsilon) f_1(x_n) + \varepsilon f_2(x_n) \quad (8)$$

where, $\varepsilon$ is the strength of the coupling $(0 < \varepsilon < 1)$, and the functions $f_1$ and $f_2$ are the chaotic trigonometric maps.

## 3. Chaotic Trigonometric Haar Wavelet Transform

In this section, we first introduce a new perspective of Haar wavelet transform and then using the chaotic trigonometric maps, we obtain chaotic trigonometric Haar wavelet transform.

- New perspective of Haar wavelet transform

As we know, the scaling function coefficients of Haar wavelet are constant values [4, 5, 6] ($p_0 = p_1 = 1$). Hence, the Haar wavelet transform is used less in cryptography. Researchers usually use of the Haar wavelet in cryptography based on the hybrid methods [25, 26, 27]. Here, we propose a new method which can improve this problem. First, we introduce a new perspective of Haar wavelet transform using of destabilized the scaling function and then the wavelet function as following method:
First, considering scaling function of Haar wavelet (Eq. 1) that changed as a sloping step function:

$$\phi(x) = \begin{cases} \lambda\left(x - \frac{1}{2}\right) + 1 & 0 \leq x < 1 \\ 0 & other\ wise \end{cases} \quad (9)$$

Where $\lambda$ is line slop and $-2 \leq \lambda \leq 2$. In result, the Haar wavelet is the function:

$$\psi(x) = \begin{cases} (\frac{\lambda^2}{24} + \frac{\lambda}{4} + 1)(2\lambda x - \frac{\lambda}{2} + 1) & 0 \le x < \frac{1}{2} \\ -(\frac{\lambda^2}{24} - \frac{\lambda}{4} + 1)(2\lambda x - \frac{3\lambda}{2} + 1) & \frac{1}{2} \le x < 1 \\ 0 & otherwise \end{cases}$$
(10)

With regard to a more comprehensive mathematical definition [4, 5], the scaling function and wavelet function of the new perspective of Haar wavelet transform are as following:

$$p_0 = \frac{\lambda^2}{24} - \frac{\lambda}{4} + 1 \,, \ p_1 = \frac{\lambda^2}{24} + \frac{\lambda}{4} + 1 \qquad (11)$$

$$\phi(x) = (\frac{\lambda^2}{24} - \frac{\lambda}{4} + 1)\phi(2x) + (\frac{\lambda^2}{24} + \frac{\lambda}{4} + 1)\phi(2x - 1)$$

$$\psi(x) = (\frac{\lambda^2}{24} + \frac{\lambda}{4} + 1)\phi(2x) - (\frac{\lambda^2}{24} - \frac{\lambda}{4} + 1)\phi(2x - 1)$$

that if $\lambda = 0$, this functions are the Haar wavelet functions. The graphs of the new scaling function and function of wavelet are given in Fig.3.

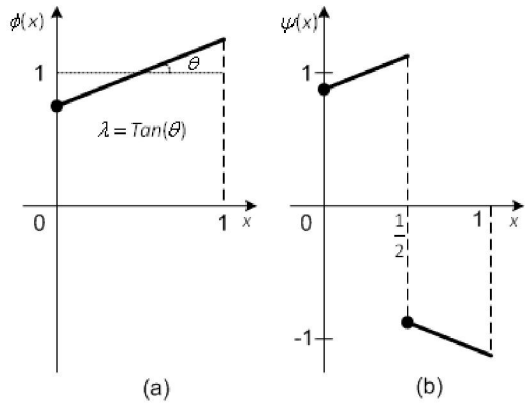

Fig.3. Graph of $\phi(x)$ and $\psi(x)$ of a new perspective of Haar wavelet.

- Basic properties of the new perspective Haar wavelet

The following concepts are basic properties in wavelets which will study in here.

1) The mean of the new perspective Haar wavelet is linear function of $\lambda$ i.e.

$$mean(\lambda) = \int_{-\infty}^{\infty} \psi(x)\,dx = 0.25\,\lambda\,.$$

This property shows that the $\lambda$ can be used in cryptography as a key.

2) The unit length of the new perspective Haar wavelet is one i.e. $\int_{-\infty}^{\infty} \phi(x)dx = 1$.

The new perspective Haar wavelet in one dimension can be shown with using the following example:

Consider the function shown in Fig. 4(a). We can approximate this function using translates of the new perspective Haar scaling function $\phi(x)$. The approximation is shown in Fig. 4(b). If we call this approximation $\phi_f^{(0)}(x, \lambda)$, then

$$\phi_f^{(0)}(x, \lambda) = \sum_k c_{0,k}(\lambda)\phi_k(x)$$

Where

$$c_{0,k}(\lambda) = \int_k^{k+1} f(x)\,\phi_k(x)\,dx$$

We can obtain a more refined approximation, or an approximation at a higher resolution, $\phi_f^{(1)}(x, \lambda)$, shown in Fig. 4(c), if we use the set $\{\phi_{1,k}(x)\}$:

$$\phi_f^{(1)}(x, \lambda) = \sum_k c_{1,k}(\lambda)\,\phi_{1,k}(x)$$

Continuing in this method (Fig. 4(c)), we can obtain higher and higher resolution approximations of $f(x)$ with

$$\phi_f^{(m)}(x, \lambda) = \sum_k c_{m,k}(\lambda)\,\phi_{m,k}(x).$$

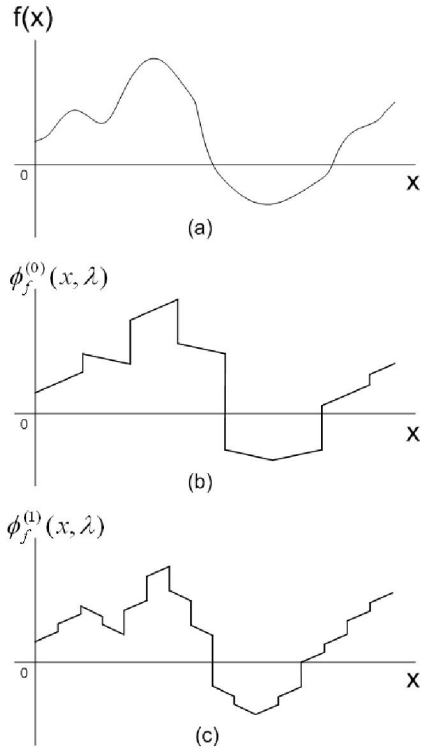

Fig. 4. A sample function and approximations of the function.

The new perspective Haar wavelet in two dimensions is obtained as similar to above mention. An example of a $4 \times 4$ new perspective Haar transformation matrix is shown in Eq. (12):

$$H_1 = \begin{vmatrix} \widetilde{p_0} & \widetilde{p_1} & 0 & 0 \\ 0 & 0 & \widetilde{p_0} & \widetilde{p_1} \\ \widetilde{p_1} & -\widetilde{p_0} & 0 & 0 \\ 0 & 0 & \widetilde{p_1} & -\widetilde{p_0} \end{vmatrix} \text{ and}$$

$$H_2 = \begin{vmatrix} \widetilde{p_0} & \widetilde{p_1} & 0 & 0 \\ \widetilde{p_1} & -\widetilde{p_0} & 0 & 0 \\ 0 & 0 & 1 & 0 \\ 0 & 0 & 0 & 1 \end{vmatrix}$$

that $H_1$ and $H_2$ are first approximation matrix and Second approximation matrix, respectively. Thus,

$$H_{4\times 4} = H_2 H_1 = \begin{vmatrix} (\widetilde{p_0})^2 & \widetilde{p_0}\widetilde{p_1} & \widetilde{p_1}\widetilde{p_0} & (\widetilde{p_1})^2 \\ \widetilde{p_1}\widetilde{p_0} & (\widetilde{p_1})^2 & -(\widetilde{p_0})^2 & -\widetilde{p_0}\widetilde{p_1} \\ \widetilde{p_1} & -\widetilde{p_0} & 0 & 0 \\ 0 & 0 & \widetilde{p_1} & -\widetilde{p_0} \end{vmatrix}$$

(12)

where the $\widetilde{p_l} = \frac{p_l}{\sqrt{2}}$ coefficients are scaling function coefficients. In result, we have a Haar wavelet transform with variable scaling function coefficients.

- Chaotic Trigonometric Haar wavelet transform

With regard to above mentions, we introduce chaotic trigonometric Haar wavelet transform with variable scaling function coefficients which is a new method in cryptography. To avoid adding content, we focus on the chaotic trigonometric Haar wavelet in two-dimension.

We use the chaotic trigonometric coupled maps to generate $\lambda$. Note that for any scaling function coefficient $(\widetilde{p_l})$ could be generated a separate $\lambda$. In result, we have a chaotic trigonometric Haar transformation matrix from variable scaling function coefficients$(\widetilde{p_l})$. A $4 \times 4$ chaotic trigonometric perspective Haar transformation matrix is as following,

$$H_1 = \begin{vmatrix} \widetilde{p_0}(\lambda_1) & \widetilde{p_1}(\lambda_2) & 0 & 0 \\ 0 & 0 & \widetilde{p_0}(\lambda_3) & \widetilde{p_1}(\lambda_4) \\ \widetilde{p_1}(\lambda_5) & -\widetilde{p_0}(\lambda_6) & 0 & 0 \\ 0 & 0 & \widetilde{p_1}(\lambda_7) & -\widetilde{p_0}(\lambda_8) \end{vmatrix} \text{ and}$$

$$H_2 = \begin{vmatrix} \widetilde{p_0}(\lambda_9) & \widetilde{p_1}(\lambda_{10}) & 0 & 0 \\ \widetilde{p_1}(\lambda_{11}) & -\widetilde{p_0}(\lambda_{12}) & 0 & 0 \\ 0 & 0 & 1 & 0 \\ 0 & 0 & 0 & 1 \end{vmatrix}$$

that $H_1$ and $H_2$ are first approximation matrix and Second approximation matrix, respectively. Then,

$$H_{4\times 4} = H_2 H_1 =$$

$$\begin{vmatrix} \widetilde{p_0}(\lambda_9)\widetilde{p_0}(\lambda_1) & \widetilde{p_0}(\lambda_9)\widetilde{p_1}(\lambda_2) & \widetilde{p_1}(\lambda_{10})\widetilde{p_0}(\lambda_3) & \widetilde{p_1}(\lambda_{10})\widetilde{p_1}(\lambda_4) \\ \widetilde{p_1}(\lambda_{11})\widetilde{p_0}(\lambda_1) & \widetilde{p_1}(\lambda_{11})\widetilde{p_1}(\lambda_2) & -\widetilde{p_0}(\lambda_{12})\widetilde{p_0}(\lambda_3) & -\widetilde{p_0}(\lambda_{12})\widetilde{p_1}(\lambda_4) \\ \widetilde{p_1}(\lambda_5) & -\widetilde{p_0}(\lambda_6) & 0 & 0 \\ 0 & 0 & \widetilde{p_1}(\lambda_7) & -\widetilde{p_0}(\lambda_8) \end{vmatrix}$$

(13)

As an example, we obtain a $4 \times 4$ chaotic trigonometric Haar transformation matrix with proposed control parameters of the chaotic trigonometric coupled maps (Eq.8) as:

$\{x = 0.2, N_1 = 3, N_2 = 4, a_1 = 2, a_2 = 2.5 \text{ and } \varepsilon = 0.4\}$.

Then,

$$\begin{cases} \lambda_1 = 1.469 & \lambda_2 = -0.351 & \lambda_3 = -0.075 & \lambda_4 = 0.027 \\ \lambda_5 = -0.070 & \lambda_6 = 0.033 & \lambda_7 = -0.028 & \lambda_8 = 0.156 \\ \lambda_9 = 0.674 & \lambda_{10} = 0.147 & \lambda_{11} = 0.570 & \lambda_{12} = 0.834 \end{cases}$$

In result, we have:

$$H_{4\times 4} = \begin{bmatrix} 0.614 & 0.780 & 1.057 & 1.044 \\ 0.835 & 1.061 & -0.836 & -0.826 \\ 0.983 & -0.992 & 0 & 0 \\ 0 & 0 & 0.993 & -0.962 \end{bmatrix}$$

We can develop this transformation matrix to a $n \times n$ chaotic trigonometric Haar transformation matrix.

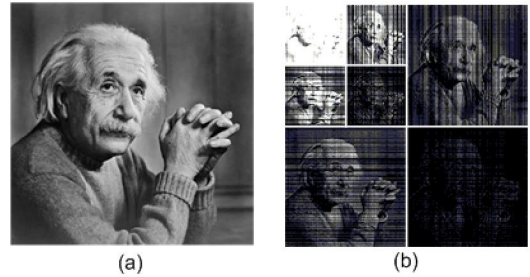

Fig.5. (a) The plain image ''Einstein.bmp''. (b) Two levels chaotic trigonometric Haar wavelet decomposition of the plain image ''Einstein.bmp''.

Here, we are done a chaotic trigonometric Haar wavelet transformation. With this method, brightness of the sub-images has randomly changed in horizontal and vertical (see Fig. 5).

## 4. Proposed algorithm

In this section, we introduce the encryption algorithm based on the chaotic trigonometric Haar wavelet transforms (CTH) and spiral swapping.

First, the CTH is used to compute the approximation coefficients matrix LL$_1$, and details coefficients matrices LH$_1$, HL$_1$, and HH$_1$ of the plain image in the first wavelet decomposition. Next, the 2nd level of the wavelet decomposition is done to generate LL$_2$, LH$_2$, HL$_2$, and HH$_2$ with new control parameters. In next step, the values of the LL$_2$ matrix involves as spiral swapping with values of the LH$_2$, HL$_2$, and HH$_2$ matrices as following diagram (see Fig. 6):

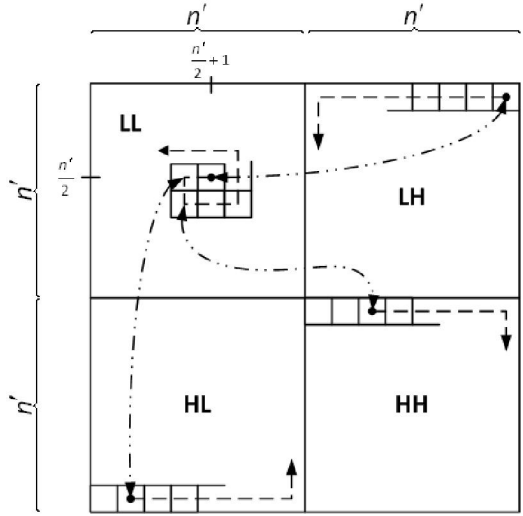

Fig.6. diagram of spiral swapping method.

As shown in the diagram, the spiral swapping method swaps $\left(\frac{\acute{n}}{2},\frac{\acute{n}}{2}+1\right)$ pixel of the LL$_2$ with $(1,\acute{n})$ pixel of the LH$_2$ and $\left(\frac{\acute{n}}{2},\frac{\acute{n}}{2}\right)$ pixel of the LL$_2$ with $(\acute{n},2)$ pixel of the HL$_2$ and also $\left(\frac{\acute{n}}{2}+1,\frac{\acute{n}}{2}\right)$ pixel of the LL$_2$ with $(1,3)$ pixel of the HH$_2$ and then swaps $\left(\frac{\acute{n}}{2}+1,\frac{\acute{n}}{2}+1\right)$ pixel of the LL$_2$ with $(1,\acute{n}-3)$ pixel of the LH$_2$ and $\left(\frac{\acute{n}}{2}+1,\frac{\acute{n}}{2}+2\right)$ pixel of the LL$_2$ with $(\acute{n},5)$ pixel of the HL$_2$ and also $\left(\frac{\acute{n}}{2},\frac{\acute{n}}{2}+2\right)$ pixel of the LL$_2$ with $(1,6)$ pixel of the HH$_2$. These steps would be continuing until $(\acute{n},\acute{n})$ pixel of the LL$_2$ and so on. The spiral swapping method will cause of brightness of the sub-images distribute as more uniform in the whole image. This step is repeated again on the 1st level of wavelet decomposition, i.e. using the spiral swapping method swap the values of the LL$_1$ matrix with those of the LH$_1$, HL$_1$, and HH$_1$. In result, a chaotic trigonometric Haar image (F) is produced with the 2-levels inverse chaotic trigonometric Haar wavelet transforms which each level has the new control parameters.

In the last step, an encrypted image is produced from combination the chaotic trigonometric Haar image and the plain image as following [31]:

$$E_{i\times j} = F_{i\times j} \; XOR \; M_{i\times j}$$

Where M is an $n \times n$ image matrix, F is a $n \times n$ chaotic trigonometric Haar image matrix, and E is an $n \times n$ encrypted image matrix.

In result, an encrypted image is obtained by this algorithm that due to having many control parameters resists any security and statistical attacks (Fig. 7). The decryption process is almost the same as the encryption but with reverse steps.

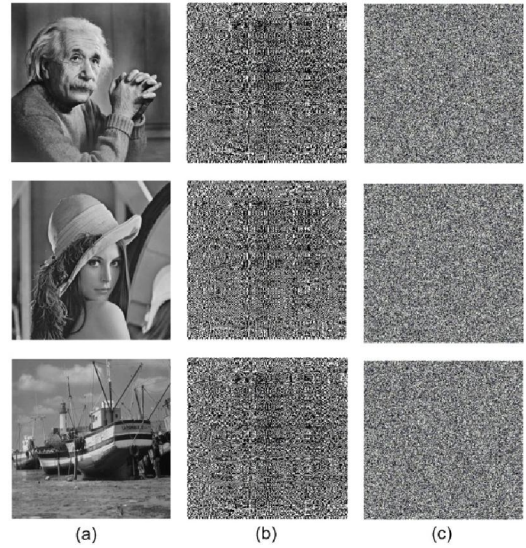

Fig.7. The figures "Einstein", "Lena", and "Boat" present the plain images (a), the chaotic trigonometric Haar images (b) encrypted images(c).

## 5. Experimental results and statistical analysis

Nowadays, security is most important part of an image encryption. Hence, a complete analysis is made on the security of the encrypted images of proposed method. We have tried based on statistical analysis to explain that the encrypted images are sufficiently secure against various cryptographically attacks.

- Key space analysis

One of the methods to check security is the key space analysis. The key space analysis is a mathematical method to obtain size of key space. If it is not large enough, the attackers may guess the image with brute-force attack [6, 28]. As we know, the total number of different keys was used in the cryptography is the Key space size. In this method, there are six control parameters at least in the interval 0 to 1 that were used four times in chaotic trigonometric Haar wavelet transforms. The Fig. 8, show the key space size the proportion to the precision of numbers.

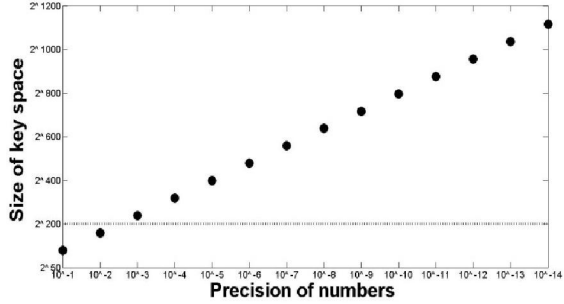

Fig. 8. The size of key space to the precision of numbers.

Hence, If the precision be rather than $10^{-3}$, the size of key spaces for control parameters of the method are much more than $2^{200}$. These sizes are large enough to defeat brute-force by any super computer today.

- Histogram

The grey scale histograms are the mathematical tools to show pixel's brightness distribution of the images. The grey scale histograms are given in Figs. 9. The grey scale histogram of the encrypted image is significantly different from the grey scale histogram of the plain image. Whatever the brightness distribution of grey scale of the encrypted image is uniformly, the encrypted image is completely encrypted and the possibility of breaking of the encrypted image is much less than exponential distribution [25]. The Fig. 9(b), show uniformity in brightness distribution of grey scale of the encrypted images. Hence, it does not provide any useful information to perform any statistical analysis attack on the encrypted image. In addition, the average pixel intensity for plain image and encrypted image are 98.92, and 127.13, respectively.

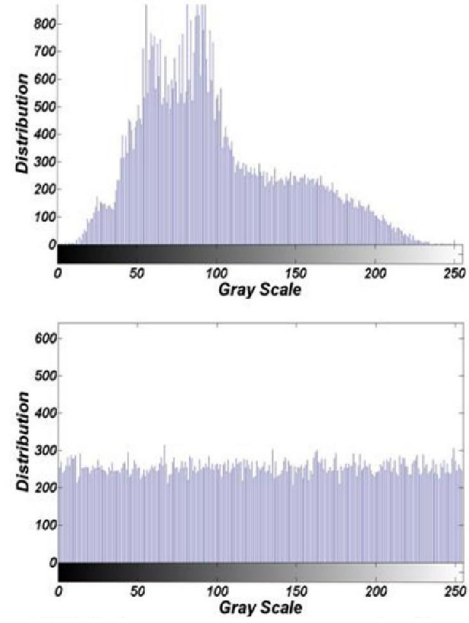

Fig.9. Histograms of plain and encrypted images.

- Normalized Information Entropy

The entropy is the most outstanding characteristic of the randomness [28]. Information entropy is a mathematical tool to measure the amount of disorder of data communication and storage that founded by Claude E. Shannon in 1949 [29]. Suppose p is the number of different values that the pixels of the encrypted image can have. Suppose $N_i$ is the amount of pixels of the encrypted image that values of the $i$ are $0, 1, \ldots, p-1$. and also, N is the total amount of pixels of the encrypted image. The entropy of the encrypted image is defined as

$$H = \sum_{i=0}^{p-1} \frac{N_i}{N} \log_2 \frac{N}{N_i}$$

which is in the interval $[0, \log_2 p]$. Normalized information entropy is a method to scale the information entropy. Hence, the normalized information entropy of the encrypted image is defined as

$$\bar{H} = \frac{\sum_{i=0}^{p-1} \frac{N_i}{N} \log_2 \frac{N}{N_i}}{\log_2 p}$$

which is in the interval [0,1]. If encrypted pixels of the intensity were equiprobable, the normalized entropy should be equal to one. In our simulations, the normalized entropy has assumed values close to one, ranging from 0.9982 to 0.9999. Therefore, our encrypted images are close to a random source with equiprobable values. In result, the proposed algorithm is secure against the entropy attack. Such result agrees with the uniformity shown in the histograms of Fig. 9.

- Correlation Coefficient analysis

In order to show the amounts of uncorrelation between pixels of encrypted image, the statistical analysis performs on the encrypted image. The correlation coefficient analysis shows the correlation between two adjacent pixels in plain image and encrypted image. We randomly select 2000 pairs of two-adjacent pixels (in vertical, horizontal, and diagonal direction) from plain image and encrypted image. We calculate the correlation coefficients based on the following two equations, respectively (see Table 1) [30, 33]:

$$Cov(x,y) = \frac{1}{N}\sum_{i=1}^{N}(x_i - E(x))(y_i - E(y))$$

$$r_{xy} = \frac{Cov(x,y)}{(D(x))^{\frac{1}{2}}(D(y))^{\frac{1}{2}}}$$

where

$$E(x) = \frac{1}{N}\sum_{i=1}^{N}(x_i), \quad D(x) = \frac{1}{N}\sum_{i=1}^{N}(x_i - E(x))^2$$

That E(x) is the estimation of mathematical expectations of x, and D(x) is the estimation of variance of x, and also Cov(x,y) is the estimation of covariance between x and y. In addition, x and y are grey scale values of two adjacent pixels in the image [30, 33].

Table. 1. Correlation coefficients of two adjacent pixels in the plain image and the encrypted images.

| Direction | Plain Image | | | Encrypted image | | |
|---|---|---|---|---|---|---|
| | Einstein | Lena | Boat | Einstein | Lena | Boat |
| Horizontal | 0.934 | 0.914 | 0.905 | 0.0015 | 0.0020 | 0.0088 |
| Vertical | 0.963 | 0.922 | 0.955 | 0.0038 | 0.0049 | 0.0138 |
| Diagonal | 0.940 | 0.949 | 0.945 | 0.0046 | 0.0057 | 0.0076 |

- Differential attack

In an attempt to derive the key, hackers try to find out a relationship between the plain image and the encrypted image, based on studying how differences in the input which can affect the resultant difference in the output [31]. Hackers observe the change of the encrypted image using a slight change such as modifying one pixel of the plain image [32, 34]. To investigate the influence of one pixel change on the whole encrypted image by the proposed algorithm, two common measures are used:

First, the number of pixels change rate (NPCR) that stands for the number of pixels change rate while one pixel of plain image is changed. Second, the unified average changing intensity (UACI) that measures the average intensity of differences between the plain image and encrypted image. The NPCR and The UACI are used to test the influence of one pixel change on the whole of the image encrypted and can be defined as following [33]:

$$NPCR = \frac{\sum_{i,j} D(i,j)}{w \times h} \times 100\%$$

$$UACI = \frac{1}{w \times h}\left[\sum_{i,j}\frac{C_1(i,j) - C_2(i,j)}{255}\right] \times 100\%$$

that w and h are the width and height of $C_1$ or $C_2$. Here, the $C_1$ and $C_2$ are two encrypted images, whose corresponding plain images have only one pixel difference and also have the same size. The $C_1(i,j)$ and $C_2(i,j)$ are grey-scale values of the pixels at grid $(i,j)$. The $D(i,j)$ determined by $C_1(i,j)$ and $C_2(i,j)$. If $C_1(i,j) = C_2(i,j)$, in result, $D(i,j) = 1$; otherwise, $D(i,j) = 0$. We have took some tests on the proposed algorithm (256 grey scale image of size $256 \times 256$) to find out the extent of change produced by one pixel change in the plain image (see Table 2). The results demonstrate that the proposed algorithm can survive differential attack.

Table. 2. Results of the differential attack in the encrypted images (Minimum, maximum and average NPCR and UACI).

| Tests | Einstein | Lena | Boat |
|---|---|---|---|
| **NPCR%** | | | |
| min: | 98.885 | 99.165 | 98.429 |
| max: | 99.125 | 99.265 | 98.695 |
| average: | 99.005 | 99.215 | 98.562 |
| **UACI%** | | | |
| min: | 24.011 | 29.188 | 37.508 |
| max: | 24.245 | 29.516 | 37.658 |
| average: | 24.128 | 29.352 | 37.583 |

## 6. Conclusion

We have introduced a new perspective of Haar wavelet transform. The new perspective of Haar wavelet transform can be used in image compression and image encryption. Using the chaotic trigonometric maps, we have introduced the chaotic trigonometric Haar wavelet transform. The algorithm was proposed to make encrypted image. The results have shown this method could be used in image encryption. We suggest the use of the new perspective of Haar wavelet transform to signal processing and image processing and also similar work.